\documentclass[lettersize,journal]{IEEEtran}
\usepackage{amsmath,amsfonts}
\usepackage{algorithmic}
\usepackage{algorithm}
\usepackage{array}
\usepackage[caption=false,font=normalsize,labelfont=sf,textfont=sf]{subfig}
\usepackage{textcomp}
\usepackage{stfloats}
\usepackage{url}
\usepackage{verbatim}
\usepackage{graphicx}
\usepackage{cite}
\usepackage{tikz}

\usepackage{verbatim}
\usepackage{color}
\usepackage{graphicx}
\begin{document}
\title{Synergetic Empowerment: Wireless Communications Meets Embodied Intelligence}
\author{\IEEEauthorblockN{Hongtao Liang, \emph{Graduate Student Member}, \emph{IEEE}, Yihe Diao, Yuhang Wu, \emph{Member}, \emph{IEEE}, \\ 
Fuhui Zhou, \emph{Senior Member}, \emph{IEEE}, and 
Qihui Wu, \emph{Fellow}, \emph{IEEE}}
\thanks{
© 2026 IEEE. Personal use of this material is permitted. Permission from IEEE must be obtained for all other uses, in any current or future media, including reprinting/republishing this material for advertising or promotional purposes, creating new collective works, for resale or redistribution to servers or lists, or reuse of any copyrighted component of this work in other works.

This work was supported in part by the National Natural Science Foundation of China under Grants 62231015, 
in part by the Yangtze River Delta Science and Technology Innovation Community Joint Research (Basic Research) Project under Grant 2024CSJZN00300, 
in part by the Basic Reserch Projects of Stabilizing Support for Specialty Disciplines under Grant ILF240041A24, 
in part by the Major Achievements Cultivation Project under Grant NC2025010, 
in part by the Fundamental Research Funds for the Central Universities under Grant NQ2025005, 
and in part by the Interdisciplinary Innovation Fund for Doctoral Students of Nanjing University of Aeronautics and Astronautics under Grant KXKCXJJ202607. (Corresponding authors: Fuhui Zhou.)

H. Liang, Y. Diao and Q. Wu are with the College of Electronic and Information Engineering, Nanjing University of Aeronautics and Astronautics, Nanjing, 210000, P. R. China.  (e-mail: ceie.lht@nuaa.edu.cn, diaoyihe@nuaa.edu.cn, wuqihui2014@sina.com).

Y. Wu and F. Zhou are with the College of Artificial Intelligence, Nanjing University of Aeronautics and Astronautics, Nanjing, 210000, P. R. China. (e-mail: may\_wyh@nuaa.edu.cn, zhoufuhui@ieee.org).
}
}
\maketitle
\begin{abstract}
Wireless communication is evolving into an agent era, where large-scale agents with inherent embodied intelligence are not just users but active participants. The perfect combination of wireless communication and embodied intelligence can achieve a synergetic empowerment and greatly facilitate the development of agent communication. An overview of this synergetic empowerment is presented, framing it as a co-evolutionary process that transforms wireless communication from a simple utility into the digital nervous system of a collective intelligence, while simultaneously elevating isolated agents into a unified superorganism with emergent capabilities far exceeding individual contributions. Moreover, we elaborate how embodied intelligence and wireless communication mutually benefit each other through the lens of the perception-cognition-execution (PCE) loop, revealing a fundamental duality where each PCE stage both challenges network capacity and creates unprecedented opportunities for system-wide optimization. Furthermore, critical open issues and future research directions are identified.
\end{abstract}


\section{Introduction}
\IEEEPARstart{W}{ireless} communication is evolving into the agent era, marking a fundamental shift from connecting passive information endpoints to enabling massive-scale agent collaboration. 
Unlike traditional devices, these agents, from autonomous vehicles to industrial robots, possess embodied intelligence, defined as a physically embodied system that perceives, reasons, and acts through direct environmental interaction while continuously refining its intelligence from experiential feedback \cite{embodied_intelligence}.
The scale of this transformation is unprecedented. The projections for 2030 estimate that the number of connected IoT devices will reach 125 billion, while monthly global mobile traffic is expected to increase to over 5000 exabytes, representing an 80-fold increase from 2020 \cite{Nguyen2022}.  
More critically, a growing portion of these devices are embodied agents that require real-time coordination for complex collective tasks. However, traditional static network architectures cannot support the highly dynamic spatial and temporal demands of these billions of moving agents. To manage this unprecedented scale, the network needs to actively harness the physical mobility and perception of these agents to dynamically optimize its own topology. Concurrently, these agents rely on the network to transcend their onboard computational constraints. This unavoidable mutual dependence necessitates a new synergetic empowerment perspective \cite{LLMMultiAgent}.

At the core of this synergetic empowerment is the tightly coupled perception-cognition-execution (PCE) loop of embodied intelligence, where perception drives execution and execution generates new perceptions. 
Autonomous navigation, precise control, and payload communication inherently rely on reliable wireless connectivity \cite{Bai2023}. This continuous cycle imposes extreme throughput and reliability demands that fundamentally reshape the function of wireless communication \cite{Nguyen2022}. The network is expected to evolve from providing best-effort delivery for passive endpoints to guaranteeing deterministic, closed-loop control for mobile agents. This fundamental shift challenges architectures designed for elastic traffic patterns. Yet these same PCE-driven agents also offer unprecedented optimization opportunities. 
Unlike passive endpoints, they are active participants whose mobility enables physical-layer network reconfiguration, and whose sensory capabilities provide high-fidelity environmental mapping, thereby transforming potential network strain into network intelligence.

\begin{figure*}[t]
	\centering
        \includegraphics[width=17cm, trim=18 18 18 18,clip]{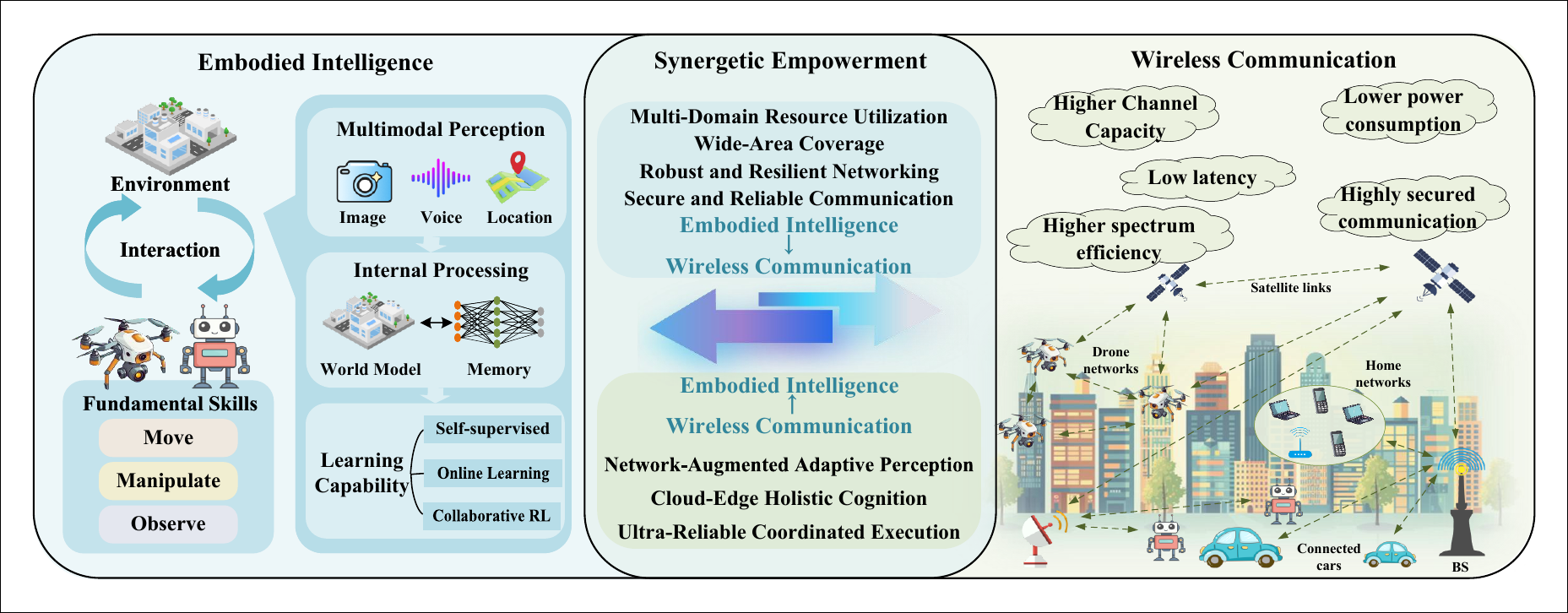}
        \caption{The synergetic empowerment between embodied intelligence and wireless communication, illustrating the core components and their mutual enhancement mechanisms.}
	\label{fig.1}
\end{figure*}

This transformation unlocks synergetic empowerment through continuous co-evolution. 
Wireless communication serves as the digital nervous system interconnecting embodied agents by relaying sensory data, cognitive outputs, and control commands, sustaining their PCE loops and augmenting cognitive capabilities \cite{EdgeAI6G}.
This network-centric framework fuses disparate sensory data from multiple agents to build holistic situational awareness, transcending the capabilities of a single entity. In turn, embodied intelligence endows the wireless communication system with physical-world intelligence through the inherent capacity of agents to sense the radio environment, make autonomous decisions, and physically alter network topology for unprecedented adaptability. 
This mutual empowerment creates a self-reinforcing feedback loop, visually summarized in Fig. \ref{fig.1}, where enhanced connectivity, collective perception, and network optimization continuously reinforce each other.

Despite the profound potential of this synergetic relationship, dedicated research remains in its infancy. Existing surveys in embodied intelligence \cite{embodied_intelligence} provide comprehensive taxonomies for the intra-agent control loop but do not extend to inter-agent collaborative dynamics mediated by communication networks. Multi-agent coordination studies \cite{ConsensusReview2022} focus primarily on consensus algorithms while treating the underlying wireless communication system as a given constraint rather than a co-evolving partner. Conversely, the communications community \cite{5G-A/6G,Bai2023} details AI-native radio techniques and network optimization but predominantly frames agents as complex service endpoints, overlooking their potential as active network constituents capable of physical reconfiguration. To bridge this gap, this article articulates a conceptual foundation for the co-evolution of embodied intelligence and wireless communication, centered on a unified cyber-physical system where the PCE loop deeply intertwines with network capabilities. Guided by this integrated view, we survey key enabling technologies, identify critical open challenges, and outline future research directions.

The remainder of this article is organized as follows. Section II introduces embodied intelligence. Sections III and IV examine the bidirectional empowerment, and Section V identifies open challenges.

\section{Embodied Intelligence}
\begin{figure*}[t]
    \centering
    \includegraphics[width=16.5cm, trim=18 18 18 18,clip]{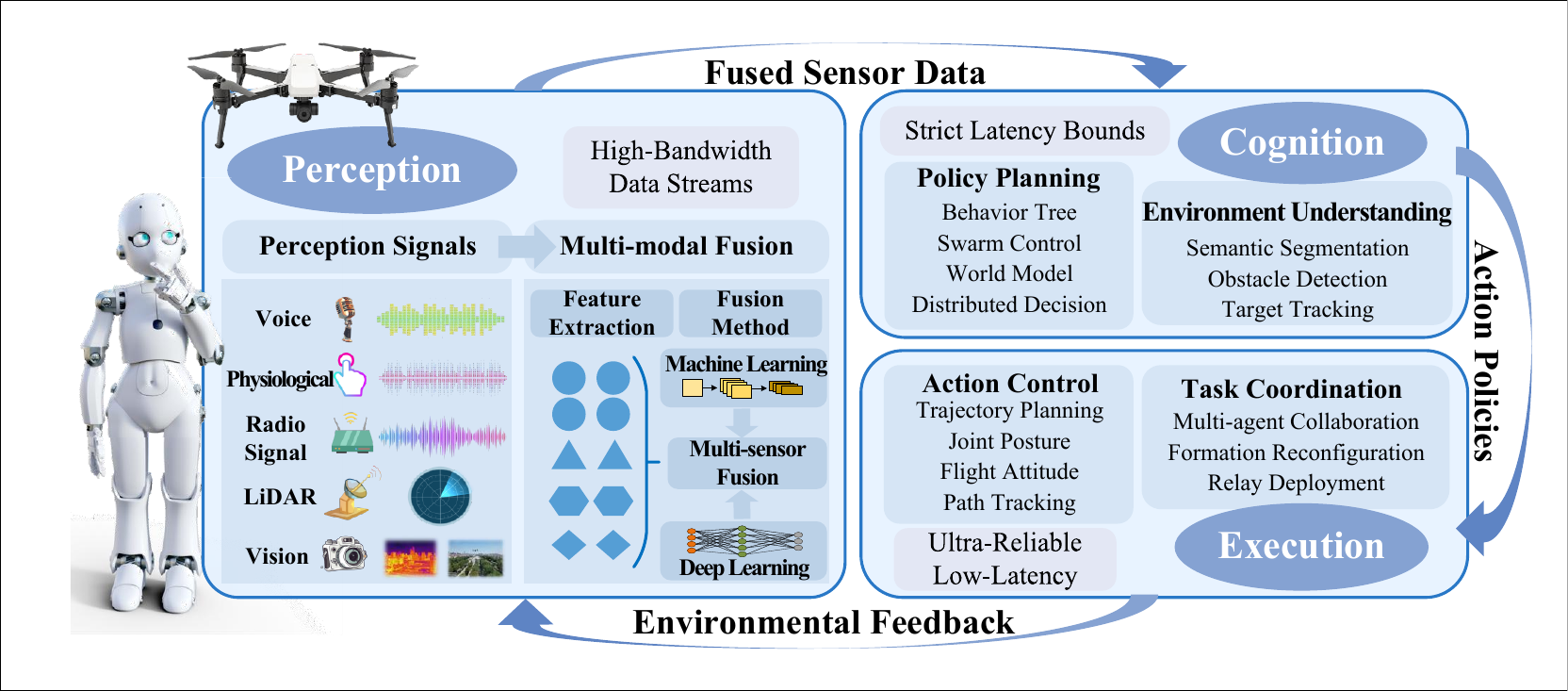}
    \caption{A conceptual framework of embodied intelligence that integrates perception, cognition, and execution. }
    \label{fig.2}
\end{figure*}

The foundational synergy between embodied intelligence and wireless communication is forged within the operational core of the agent itself. 
Embodied intelligence operates through the tight PCE loop, where perception, cognition, and execution overlap in a pipelined fashion with continuous cross-stage feedback, as illustrated in Fig.~\ref{fig.2}.
This entire process is not self-contained but represents a continuous, dynamic interface with the ambient wireless communication system \cite{embodied_intelligence}. 
Through this interface, the physical actions of the agent and the digital state of the network become deeply interdependent. This interdependence gives rise to a fundamental duality, which is revealed in each stage of the loop by imposing extreme demands on network performance while unlocking transformative opportunities for network optimization.

\subsection{Environment Perception}

Perception in embodied intelligence transcends passive data collection. It is an active process of physical exploration where the interaction between an agent and its environment generates a continuous stream of high-dimensional, multi-modal sensory data \cite{embodied_intelligence}. 
This operational model starkly contrasts with traditional static sensing. Embodied perception from mobile agents generates high-dimensional and multi-modal data streams, such as continuous high-resolution video and dense 3D point clouds. This creates a dynamic and bandwidth-intensive uplink load that challenges conventional wireless communication system architectures.
Therefore, a high-performance communication link is not merely an enabler but an intrinsic component of the advanced perceptual process.

The data stream generated by this process creates a fundamental duality for the network. On one hand, the stream is a massive communication payload, representing a significant transport burden that demands extreme fidelity. On the other hand, this same data provides rich environmental metadata, an asset the wireless communication system can exploit for its own optimization. 
An autonomous vehicle provides a clear illustration of this duality \cite{Luo2024}. While the wireless communication system is burdened with transmitting the continuous sensor streams of the vehicle, analysis performed at the network edge can simultaneously exploit this data to identify signal-blocking structures, both static like tunnels and dynamic like large vehicles. This real-time information then enables proactive  and predictive management of the communication link.
Thus, for the network, perceptual data is transformed from a mere transport challenge into a valuable source of real-time environmental intelligence \cite{Xu2023}.

\subsection{Situation Cognition}
The central role of cognition in embodied intelligence is to transform the high-volume, low-value-density data streams from perception into low-volume, high-value streams of knowledge and intent that are typically more concise and semantically rich. This operational model fundamentally differs from the traditional distributed computing, which often involves unidirectional data transfers for offline analysis. 
Embodied cognition requires the wireless communication system to support a persistent and bidirectional dialogue among agents and the computing nodes. This continuous state synchronization demands strict latency bounds to prevent the agents from executing critical physical maneuvers based on obsolete environmental states \cite{LLMEdgeAI}.
Meaningful cognition at scale is therefore inherently a networked concept. Without a robust and reliable communication architecture as its foundation, advanced collective intelligence cannot emerge.

When this networked cognition is operationalized, it imposes a profound workload duality upon the wireless communication system. On one hand, the network provides the high-bandwidth, low-latency connectivity required to support complex distributed reasoning and state synchronization across multiple nodes. On the other hand, the network gains an unprecedented semantic opportunity by receiving and interpreting the high-level intent streams distilled by the cognitive process, enabling a shift from reactive service management to proactive, goal-oriented resource allocation \cite{Xu2023}. This duality is illustrated within a smart factory, where the network simultaneously provides ultra-reliable links for collaborative robot coordination and acts upon the resulting strategic directives to prioritize critical production lines, transforming itself from a mere information conduit into an active participant in goal-oriented reasoning.

\subsection{Mission-Oriented Execution}
Execution closes the loop by translating cognitive strategies into tangible physical actions that run concurrently with ongoing perception and cognition.
Unlike simple remote actuation with open-loop commands, embodied execution is integral to the closed loop, where every action immediately generates new perceptual feedback informing the next cycle \cite{embodied_intelligence}. For any multi-agent system, execution itself is fundamentally a communication-dependent act. Collective execution, such as synchronized movement or collaborative manipulation, is impossible without a high-reliability, low-latency network to coordinate actions. Communication is therefore the prerequisite for coherent physical action at scale.

The physical execution of tasks imposes a final, profound duality upon the network. 
The network bears the control burden of mission-critical execution streams. Although physical actuation commands typically have small payloads, they demand ultra-reliable low-latency communication (URLLC) with strict deterministic guarantees, as a single delayed packet can cause immediate physical collisions \cite{AINative6G}.
Conversely, physical action itself becomes a powerful optimization tool. The movement of an agent can dynamically reconfigure the network topology, transforming the agent from a mere user into a physical reconfiguration asset. This is powerfully demonstrated by a UAV swarm where the network first delivers perfectly synchronized flight commands for collision avoidance \cite{Bai2023}. Yet, when one UAV enters a signal-dead zone, another can reposition to act as a physical relay, restoring the link for the team. Execution thus closes the loop, transforming the agent from an endpoint controlled by the wireless communication system into an active participant that physically reshapes the wireless communication system itself.

\section{Embodied Intelligence Empowering Wireless Communication}
Embodied intelligence fundamentally redefines the relationship between an agent and the network, recasting the agent from a passive service endpoint into an active, intelligent constituent capable of co-evolving with the wireless communication system. This view inverts the traditional approach of network design, which is architected for a population of passive clients under a static, top-down management model. In this new approach, the wireless communication system is architected to leverage the distributed intelligence and physical capabilities of these agents, transforming optimization into a dynamic, bottom-up process deeply coupled with the physical world \cite{ConsensusReview2022}. 
Harnessing this symbiotic relationship enables the network to transcend conventional boundaries, though constrained by mission-communication tradeoffs and the finite onboard energy budgets of each agent.

\begin{figure}[t]
    \centering
    \includegraphics[width=8cm, trim=18 32 18 25,clip]{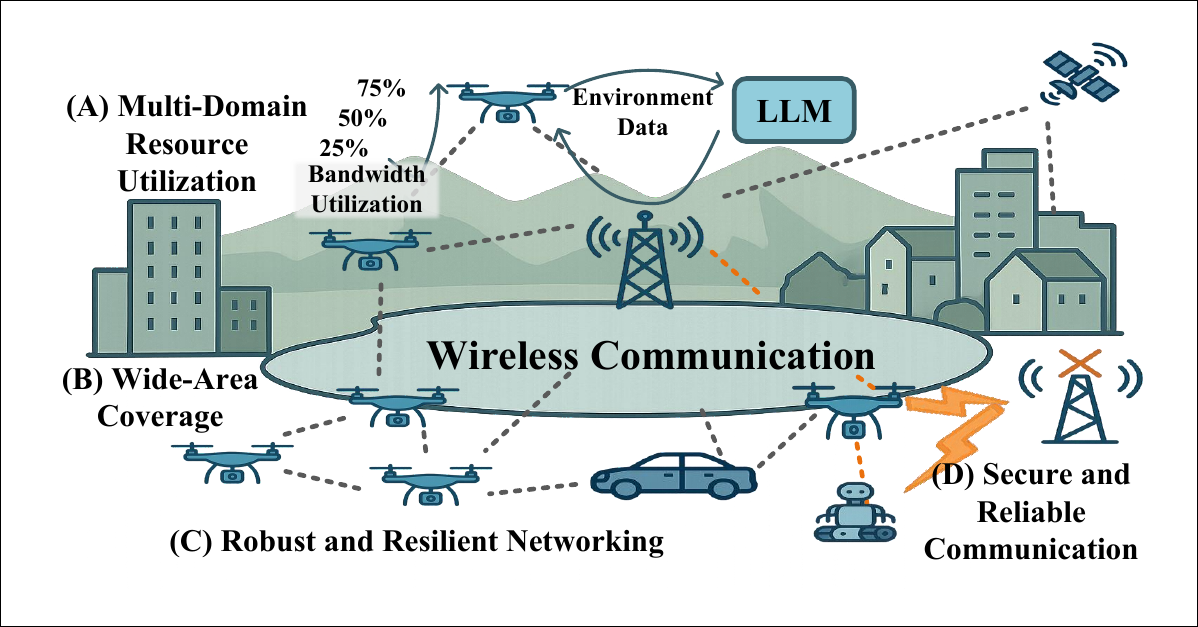}
    \caption{Embodied intelligence strengthens wireless communication by enabling multi-domain resource utilization, wide-area coverage, robust and resilient networking, and secure and reliable communication.}
    \label{Fig.3}
\end{figure}

This capacity for self-optimization is realized as agents dynamically assume differentiated roles based on real-time PCE state. As depicted in Fig.~\ref{Fig.3}, block (A) shows cognitive functions driving adaptive multi-domain resource utilization. Blocks (B) and (C) illustrate agents physically reconfiguring topology for wide-area coverage and resilient networking. Block (D) captures the closed-loop security process from threat perception to physical countermeasure execution.
The following subsections examine these key dimensions of empowerment in detail.

\subsection{Multi-Domain Resource Utilization}

Embodied intelligence reshapes multi-domain resource utilization by shifting from traditional top-down allocation to bottom-up embodied optimization. Rather than treating time, frequency, space, power, and computation as disjointed quantities assigned to passive devices, the agent considers external network resources and its own physical attributes as a unified, co-optimizable pool. Through the PCE loop, it perceives its multi-domain resource status, cognitively derives a cross-domain strategy, and executes accordingly \cite{LLMEdgeAI}.

The power of this embodied approach is evident across multiple resource domains. Consider an unmanned aerial vehicle (UAV), which can leverage its physical mobility to co-optimize the use of space and frequency resources \cite{Bai2023}. 
Upon perceiving poor channel quality, the agent cognitively evaluates real-time environmental metadata to compute optimal spatial coordinates and executes a precise flight maneuver.
Such physical repositioning can triple average user throughput, a powerful form of optimization unavailable to any static device \cite{Bai2023}.
This principle extends to the internal resources of the agent, enabling a novel trade-off between computation, energy, and time \cite{EdgeAI6G}. 
Agents can encapsulate their internal states within routine perception streams, thereby informing network-level scheduling decisions. For instance, an agent perceiving lenient deadlines can lower its CPU frequency to conserve diminishing energy.
This action sacrifices computational speed to extend operational lifetime. By integrating physical action and internal state management into the optimization loop, embodied intelligence achieves holistic, cyber-physical resource utilization that offers superior efficiency and resilience.

\subsection{Wide-Area Coverage}

Conventional coverage relies on a supply-driven model that deploys fixed base stations presuming alignment with unpredictable demand, resulting in chronic resource misalignment \cite{Zaid2023}. Embodied intelligence inverts this paradigm to be demand-driven, empowering the network to actively perceive where demand exists and physically deliver coverage as a targeted, on-demand resource.

This is illustrated by a post-disaster scenario with absent ground infrastructure. A fleet of UAVs is first deployed to perceive the demand landscape by identifying the precise locations of user signals from survivors or rescue teams \cite{LGMDigitalTwin6G}. 
This spatial and signal-strength information feeds a distributed cognitive algorithm that calculates optimal hover coordinates for maximum coverage and continuously updates routing paths to maintain mesh integrity. Subsequently, the agents physically execute these calculated trajectories, autonomously adjusting their spatial configurations to stabilize the network links against environmental disturbances and dynamic user movements.
By continuously iterating this PCE loop, network coverage becomes a living, adaptive fabric, ensuring that finite resources are directed toward points of maximum impact.

\subsection{Robust and Resilient Networking}
In high-mobility scenarios with rapidly fluctuating channels, conventional reactive methods relying on fixed signal thresholds struggle to adapt \cite{Zaid2023,AINative6G}. Embodied intelligence shifts the paradigm from reactive recovery to proactive prediction, leveraging the PCE loop to anticipate future link degradation and execute corrective actions before service quality degrades.

This proactive capability is operationalized through the PCE loop of an agent navigating a dynamic environment. Consider a high-speed train traversing a landscape with heterogeneous connectivity options, including terrestrial and satellite links \cite{5G-A/6G}. An embodied agent on the train does not merely measure the current link but actively perceives a rich dataset, including the signal quality of all nearby cells, its own velocity and trajectory, and digital terrain maps of the path ahead. 
These kinematic features and historical link statistics directly feed the predictive cognitive model.
The system calculates not just its present condition but its future state, reasoning that in five seconds the current link will be obstructed while a satellite link will become optimal. The agent executes a pre-emptive handover before service quality degrades, for example by pre-authenticating with the new link and switching at the calculated moment. By leveraging the PCE loop to translate multi-modal perception into predictive cognition and pre-emptive execution, embodied intelligence transforms a disruptive network event into an invisible optimization process, achieving robustness unattainable by reactive methods \cite{NetworkedISAC}.

\subsection{Secure and Reliable Communication}
Conventional security relies on static, perimeter-based defenses with predefined rules, ill-equipped for the dynamic topologies and physical threats inherent to embodied agent networks. Embodied intelligence replaces this with a dynamic adaptive immune system, where agents collectively perceive threats, cognize coordinated responses, and execute physical countermeasures to achieve autonomous resilience \cite{ContextSecurity6G}.

The capabilities of this adaptive immune system are enacted through the PCE loop of agents operating in a contested environment. For instance, consider a fleet of autonomous agents encountering a sophisticated physical threat, such as a targeted jammer. 
An agent on the periphery perceives abnormal link degradation. However, reliably distinguishing deliberate jamming from benign channel fading requires correlating signal anomaly patterns across spatially diverse agents. Once this collective cross-validation confirms the threat, the fleet cognizes the jammer location and formulates a topological countermeasure \cite{ConsensusReview2022}.
The fleet then executes this solution. Unaffected nodes reroute traffic away from the compromised area, while a designated agent physically repositions to serve as a relay and bridge the new communication path. By integrating physical execution into the defense loop, this fusion of cyber awareness and physical agility creates a security posture far more resilient than any static, rule-based system.

\section{Wireless Communication Empowering Embodied Intelligence}
Advanced wireless communication extends the boundaries of embodied intelligence, liberating its potential from the confines of physical embodiment. 
It transforms an ensemble of isolated, capability-limited agents into a collective superorganism, a unified entity whose coordinated actions yield emergent capabilities far exceeding individual contributions \cite{ConsensusReview2022}.
This view recasts the wireless communication system from a passive data conduit into the digital nervous system of the PCE loop, extending sensory reach to remote perspectives, augmenting cognitive capacity with cloud-based computation, and synchronizing physical actions into a collective will.
This networked substrate unlocks collective capabilities unassailable by any individual agent, though at the cost of coordination overhead and strict dependence on network reliability for physical safety.

\begin{figure}[t]
    \centering
    \includegraphics[width=8cm, trim=15 25 15 20,clip]{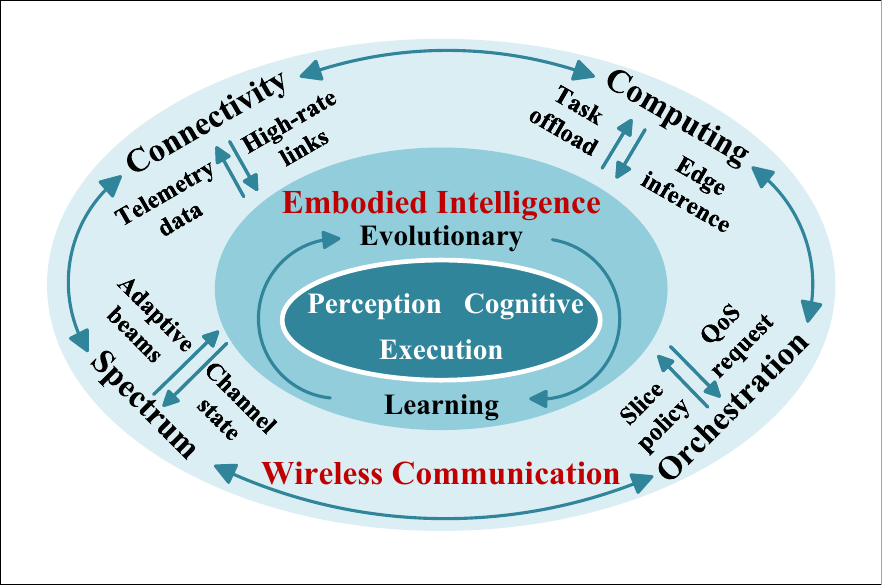}
    \caption{Wireless communication extends the capabilities of embodied agents by offering high-throughput connections, edge computing resources, and universal network access for enhanced perception, cognition, and execution.}
    \label{Fig.4}
\end{figure}

This profound empowerment is achieved through the direct augmentation of each stage within the operational PCE loop, as illustrated in Fig.~\ref{Fig.4}. 
For perception, the connectivity interface leverages sidelink and integrated sensing-communication protocols for task-oriented encoding of sensory streams. For cognition, the computing interface harnesses multi-access edge computing for semantic compression and inference, bridging onboard processors to cloud resources. Underpinning these stages, the spectrum interface employs massive multiple-input multiple-output (MIMO) beamforming driven by channel state information (CSI). For execution, the orchestration interface maps quality-of-service requirements onto dedicated URLLC network slices for knowledge-aware coordinated execution.

\subsection{Network-Augmented Adaptive Perception}
Perception in an isolated embodied agent is fundamentally limited by its physical location and onboard resources, resulting in a fragmentary understanding of the environment. Advanced wireless communication overcomes this limitation by transforming perception from a solitary, ego-centric process into collective, holistic situational awareness. Through the wireless communication system, an agent can assimilate the sensory data of its peers, fusing multiple partial viewpoints into a single, unified, and more veridical model of reality \cite{Luo2024}. This fused perspective allows the agent ensemble to overcome individual line-of-sight limitations and perceive large-scale environmental phenomena invisible to any single member, resulting in a high-fidelity world model far exceeding what any individual agent could construct alone.

The realization of this collective awareness is enabled by a network-driven PCE loop, where shared sensory data informs a superior cognitive model that in turn guides coordinated execution \cite{NetworkedISAC}. This principle is demonstrated in autonomous platooning, where ultra-reliable vehicle-to-vehicle (V2V) communication enables the perception of a distant, occluded road hazard from a lead truck, precisely time-stamped and geo-tagged, to be instantaneously disseminated throughout the platoon. The distributed cognitive system of the fleet fuses this data to cognize a shared threat and formulate a coordinated braking maneuver, which the entire platoon then executes in perfect synchrony. 
Thus, network-fused perception can extend detection range by 3.6 times and improve accuracy by 20\% compared to isolated single-agent sensing, transforming limited observers into a cohesive perceptual system \cite{Luo2024}.

\subsection{Cloud-Edge-Client Holistic Cognition}
Whereas perception furnishes the network with a torrent of digital information, the core challenge of cognition is to distill this raw data into meaningful semantic information \cite{Xu2023}.  
The onboard processors can run efficient small models within real-time bounds such as 100 ms but lack the capacity for the deep reasoning required to understand context, causality, and abstract intent \cite{EdgeAI6G}.
Advanced wireless communication facilitates semantic augmentation through a collaborative framework between cloud-hosted large language models (LLMs) and on-board small models \cite{LGMDigitalTwin6G}. The wireless communication system acts as a semantic bridge, allowing the local model of an agent to escalate situations of high semantic uncertainty to a powerful vision-language model (VLM) in the cloud for comprehensive contextual understanding and strategic guidance. 
This creates a hybrid intelligence, where client-edge co-inference compresses transmission volume from 224 KB to 33 KB while maintaining task accuracy, transforming raw data processing into semantic reasoning \cite{LLMEdgeAI}.

This semantic augmentation is enacted through a PCE loop that supports semantic escalation \cite{Xu2023}. 
The onboard model of an agent manages the high-frequency perceptual loop, continuously monitoring its own inference confidence and triggering cloud escalation only when this certainty metric falls below a task-specific threshold.
Consider a logistics robot that perceives a novel spill. Its local model can process shape and color information but cannot ascertain the semantic meaning of the substance. The agent escalates this query to a cloud-hosted multi-modal model, which references a global knowledge base to identify the substance as a corrosive chemical and prescribe a safety protocol.
Before executing this directive, the agent cross-validates it against local sensor ground truth to mitigate potential hallucinations and domain biases inherent in large models. Over time, local models refine their recognition boundaries through lightweight distillation of cloud-derived knowledge, gradually reducing escalation frequency and strengthening local autonomy without prohibitive on-device LLM training.

\subsection{Ultra-Reliable Coordinated Execution}
In the absence of a high-performance communication fabric, the execution capabilities of multi-agent systems are severely constrained. Their actions are disjointed and asynchronous, restricting them to simple, non-interactive parallel tasks where precise physical coordination is not required. Advanced wireless communication overcomes this by enabling a shift from individual actions to synchronized, emergent collective behavior \cite{LLMMultiAgent}. In this approach, the network functions as a synchronization backbone. 
Protocols like URLLC, achieving end-to-end latency as low as 10 ms, allow the physical actions of individual agents to fuse into a single, cohesive collective action, enabling the emergence of new physical capabilities that qualitatively exceed individual contributions \cite{Zaid2023}.

This emergence of collective behavior is enabled by a rapid-cycle PCE loop underpinned by URLLC. The critical importance of this principle is demonstrated when a team of construction robots collaboratively assembles a delicate, heavy structure. A shared cognitive plan governs the operation. 
During execution, the URLLC network synchronizes motor commands and shares minute force and position variations for real-time collective compensation. Crucially, if link quality degrades beyond acceptable bounds, each agent autonomously reverts to a pre-computed, locally stored safe-halt posture, preventing uncoordinated motion until connectivity is restored.
This precise synchronization of physical action is also what enables remote robotic surgery, where the delicate maneuvers of a surgeon are replicated with high fidelity and near-zero latency, and autonomous platooning, where vehicles execute seamless, coordinated maneuvers \cite{LLMMultiAgent}. Thus, the wireless communication system functions as a digital muscle fiber, binding individual actuators together to form a cohesive super-actuator capable of performing highly complex tasks requiring collective precision and exceptional strength.

\section{Open Issues and Challenges}
\begin{figure}[t]
    \centering
    \includegraphics[width=8cm, trim=15 15 15 15,clip]{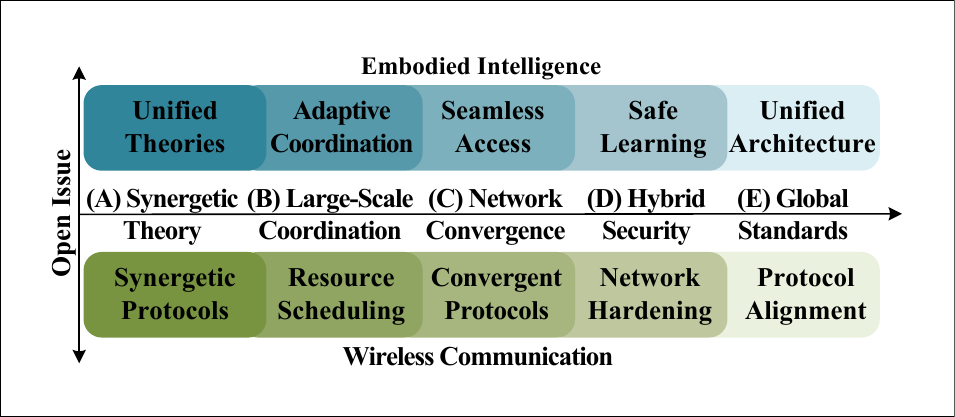}
    \caption{Overview of key open issues in embodied intelligence and wireless communication integration.}
    \label{Fig.5}
\end{figure}
While this synergy promises transformative capabilities, realizing the vision requires confronting challenges arising directly from the deep integration of cyber and physical domains, demanding breakthroughs in foundational theory, system architecture, and operational trust.
This section scrutinizes these critical open issues, categorized in Fig.~\ref{Fig.5}, spanning longer-term foundational challenges in cross-domain theory and global standardization, near-term bottlenecks in coordination scalability and network convergence, and security concerns that cut across both horizons.

\subsection{Multi-level Development of Synergetic Theory}
The foremost challenge lies in forging a unified theoretical language to describe a system where the laws of information theory and the principles of physical dynamics are no longer separate but deeply entangled. This synergy creates a system where communication protocols account for physical state, and physical actions are themselves a form of network optimization.
Promising initial methodologies include co-simulation within shared digital twins that jointly model channel dynamics and agent kinematics, yet foundational questions remain unanswered.
For instance, how do we formulate a unified metric that bridges traditional communication measures such as latency and reliability with task-level performance indicators such as perception confidence and control stability?
This is uniquely difficult because physical actions continuously reshape the wireless channel carrying subsequent control commands, coupling information-theoretic and physical dynamics in ways that demand a new mathematical framework.

\subsection{Large-Scale Multi-Agent Collaboration and Self-Organizing Networking}
The central challenge of scaling is to manage the tension between the autonomy of individual agents and the stability of the collective superorganism. As the number of agents grows into the millions, the digital nervous system that connects them faces risks of information overload, decision latency, and cascading failures. 
The core question is how to design local interaction rules that guarantee globally stable emergent behavior, which is uniquely difficult because autonomous physical movement reshapes the shared wireless topology, so local decisions can destabilize collective connectivity. Developing scalable algorithms where agents self-organize into local clusters that aggregate perception and compress inter-cluster traffic is essential for preventing per-agent communication overhead from overwhelming network capacity and latency budgets as population density grows.

\subsection{Efficient Integration of Heterogeneous Networks}
The challenge in network integration is to transform a patchwork of disparate communication technologies, such as 5G, low Earth orbit (LEO) satellites, and Wi-Fi, into a single, seamless networked substrate for the collective superorganism. 
This is uniquely demanding because a single PCE cycle simultaneously requires high-throughput uplink for perception, low-latency edge access for cognition, and ultra-reliable downlink for execution, all while the agent traverses coverage boundaries of these disparate networks. A concrete research initiative is developing an intent-driven orchestration layer that automatically translates PCE-level mission requirements into technology-agnostic resource reservations spanning terrestrial, satellite, and local links simultaneously.

\subsection{Comprehensive Security and Autonomous Protection}
The features that make this synergy powerful, namely physical embodiment and deep cyber-physical integration, also create unprecedented hybrid attack surfaces. The challenge is no longer merely to protect the cyber state of a system but to secure the learning and adaptation process of the adaptive immune system itself. 
This is uniquely severe because attacks at the cyber-physical seam simultaneously corrupt digital reasoning and physical behavior, and the distributed nature of collective adaptation amplifies even localized adversarial poisoning into system-wide vulnerability. 
A concrete research initiative is developing adversarial training pipelines that systematically expose agents to synthetic cyber-physical attacks during simulation, building verifiable resilience across communication and control layers before deployment.

\subsection{Standardization and International Collaboration}
The ultimate challenge for implementation is to establish the standards of this synergy for the agent communications. 
The task requires a new abstraction layer where agents request resources through mission-level semantic interfaces, decoupled from the heterogeneous transport technologies beneath. This is uniquely complex because the required standards need to encode not only conventional data formats but also the semantic intent of autonomous agents and real-time physical capabilities of the network, bridging two domains that have historically evolved in complete isolation. Reconciling the independently evolved protocol stacks of robotics and telecommunications demands cross-domain harmonization that jointly defines common semantic interfaces and interoperability profiles.
Without unified protocols standardizing how perception sharing, cognitive escalation, and execution coordination directives are encoded and exchanged, the ecosystem fractures into incompatible proprietary silos.

\section{Conclusions}
The synergetic empowerment between embodied intelligence and wireless communication was comprehensively reviewed in this work. Through the PCE loop, a fundamental duality was revealed at each stage. Perception data simultaneously constitutes a transport burden and a source of environmental intelligence. Cognition distills this sensory volume into semantic directives for goal-oriented resource allocation, while execution transforms agents into physical reconfiguration assets that reshape network topology. Building on this duality, agent mobility and sensing provide bottom-up network adaptability, while connectivity and edge computing fuse isolated agents into a coordinated collective with emergent capabilities exceeding individual contributions. Realizing this vision requires resolving the coupling of information-theoretic and physical dynamics, developing scalable self-organizing protocols, and building security resilient to hybrid cyber-physical threats.

\textbf{Biographies}
{\small

Hongtao Liang (Graduate Student Member, IEEE) received the B.Eng. degree in information engineering from Nanjing University of Aeronautics and Astronautics, Nanjing, China, in 2023, where he is currently pursuing the Ph.D. degree in communication and information systems with the School
of Electronic and Information Engineering. His current research interests include embodied artificial intelligence, deep learning, and deep reinforcement learning.

Yihe Diao is currently pursuing the Ph.D. degree with the School of Electronic and Information Engineering, Nanjing University of Aeronautics and Astronautics. His current research interests include deep reinforcement learning, embodied intelligence, and their applications in cognitive UAV networks, spectrum management for wireless communications networks including spectrum aggregation, intelligent spectrum sharing, and air-ground communications.

Yuhang Wu (Member, IEEE) received the M.S. degree in information and communication engineering from Nanchang University, Nanchang, China, in 2020, and the Ph.D. degree in information and communication engineering from Nanjing University of Aeronautics and Astronautics, China, in 2025. She is currently a Postdoctoral Researcher with College of Artificial Intelligence, Nanjing University of Aeronautics and Astronautics. Her research interests include convex optimization, cognitive radio, low altitude intelligent network, and wireless resource allocation.

Fuhui Zhou (Senior Member, IEEE) is currently a Full Professor with Nanjing University of Aeronautics and Astronautics, Nanjing, China, where he is also with the Key Laboratory of Dynamic Cognitive System of Electromagnetic Spectrum Space. He has been selected for one ESI hot paper and 13 ESI highly cited papers. He has published over 200 papers in internationally renowned journals and conferences in the field of communications. His research interests include cognitive radio, cognitive intelligence, knowledge graph, edge computing, and resource allocation. He has received four best paper awards at international conferences, such as IEEE GlobeCom and IEEE ICC. He was awarded as the 2021 Most Cited Chinese Researchers by Elsevier, the Stanford World’s Top 2\% Scientists, the IEEE ComSoc Asia–Pacific Outstanding Young Researcher, and the Young Elite Scientist Award of China and URSI GASS Young Scientist. He serves as an Editor for IEEE TRANSACTIONS ON COMMUNICATIONS, IEEE SYSTEMS JOURNAL, IEEE WIRELESS COMMUNICATIONS LETTERS, IEEE ACCESS, and Physical Communications.

Qihui Wu (Fellow, IEEE) received the B.S. degree in communications engineering, the M.S. and Ph.D. degrees in communications and information systemsfrom the Institute of Communications Engineering, Nanjing, China, in 1994, 1997, and 2000, respectively. From 2003 to 2005, he was a Postdoctoral Research Associate with Southeast University, Nanjing, China. From 2005 to 2007, he was an Associate Professor with the College of Communications Engineering, PLA University of Science and Technology, Nanjing, China, where he was a Full Professor from 2008 to 2016. SinceMay 2016, he has been a Full Professor with the College of Electronic and Information Engineering, Nanjing University of Aeronautics and Astronautics, Nanjing, China. From March 2011 to September 2011, he was an Advanced Visiting Scholar with the Stevens Institute of Technology, Hoboken, USA. His current research interests span the areas of wireless communications and statistical signal processing, with emphasis on system design of software defined radio, cognitive radio, and smart radio.


\begin{thebibliography}{1}
\bibliographystyle{IEEEtran}

\bibitem{embodied_intelligence}
H.~Liu, D.~Guo, and A.~Cangelosi, ``Embodied intelligence: A synergy of morphology, action, perception and learning,'' \textit{ACM Comput. Surv.}, vol.~57, Art.~186, Feb.~2025.

\bibitem{Nguyen2022}
D.~C.~Nguyen, M.~Ding, P.~N.~Pathirana, A.~Seneviratne, J.~Li, and D.~Niyato, ``6G Internet of Things: A comprehensive survey,'' \textit{IEEE Internet Things J.}, vol.~9, no.~1, pp.~359--383, Jan.~2022.

\bibitem{LLMMultiAgent} 
F.~Jiang, L.~Dong, Y.~Peng, K.~Wang, K.~Yang, C.~Pan, D.~Niyato, and O.~A.~Dobre, ``Large language model enhanced multi-agent systems for 6G communications,'' \textit{IEEE Wireless Commun.}, vol.~31, no.~3, pp.~69--75, Jun.~2024.

\bibitem{Bai2023}
Y.~Bai, H.~Zhao, X.~Zhang, Z.~Chang, R.~Jäntti, and K.~Yang, ``Toward autonomous multi-UAV wireless network: A survey of reinforcement learning-based approaches,'' \textit{IEEE Commun. Surveys Tuts.}, vol.~25, no.~4, pp.~3038--3067, Fourth Quarter 2023.

\bibitem{EdgeAI6G}
K.~B.~Letaief, Y.~Shi, J.~Lu, and J.~Lu, ``Edge Artificial Intelligence for 6G: Vision, Enabling Technologies, and Applications,'' \textit{IEEE J. Sel. Areas Commun.}, vol.~40, no.~1, pp.~5--36, Jan.~2022.

\bibitem{ConsensusReview2022} 
A.~Amirkhani and A.~H.~Barshooi, ``Consensus in multi-agent systems: A review,'' \textit{Artif. Intell. Rev.}, vol.~55, no.~5, pp.~3897--3935, Jun.~2022.

\bibitem{5G-A/6G}
Q.~Cui \textit{et al.}, ``Overview of AI and communication for 6G network: fundamentals, challenges, and future research opportunities,'' \textit{Sci. China Inf. Sci.}, vol.~68, Art.~171301, Apr.~2025.

\bibitem{Luo2024}
G.~Luo, C.~Shao, N.~Cheng, H.~Zhou, H.~Zhang, and Q.~Yuan, ``EdgeCooper: Network-aware cooperative LiDAR perception for enhanced vehicular awareness,'' \textit{IEEE J. Sel. Areas Commun.}, vol.~42, no.~1, pp.~207--222, Jan.~2024.

\bibitem{Xu2023}
J.~Xu, T.~Y.~Tung, B.~Ai, W.~Chen, Y.~Sun, and D.~Gündüz, ``Deep joint source-channel coding for semantic communications,'' \textit{IEEE Commun. Mag.}, vol.~61, no.~11, pp.~42--48, Nov.~2023.

\bibitem{LLMEdgeAI} Y.~Shen, J.~Shao, X.~Zhang, Z.~Lin, H.~Pan, D.~Li, J.~Zhang, and K.~B.~Letaief, ``Large language models empowered autonomous edge AI for connected intelligence,'' \textit{IEEE Commun. Mag.}, vol.~62, no.~10, pp.~140--146, Oct.~2024.

\bibitem{AINative6G} 
J.~Hoydis, F.~A.~Aoudia, A.~Valcarce, and H.~Viswanathan, ``Toward a 6G AI-native air interface,'' \textit{IEEE Commun. Mag.}, vol.~59, no.~5, pp.~76--81, May~2021.

\bibitem{Zaid2023}
A.~A.~Zaid, B.~E.~Y.~Belmekki, and M.-S.~Alouini, ``eVTOL communications and networking in UAM: Requirements, key enablers, and challenges,'' \textit{IEEE Commun. Mag.}, vol.~61, no.~8, pp.~154--160, Aug.~2023.

\bibitem{LGMDigitalTwin6G} 
Y.~Yang, W.~Sun, J.~He, Y.~Fu, and L.~Xu, ``Large generative model-enabled digital twin for 6G networks,'' \textit{IEEE Netw.}, vol.~39, no.~3, pp.~29--36, May~2025.

\bibitem{NetworkedISAC} 
J.~Li, X.~Shao, F.~Chen, S.~Wan, C.~Liu, Z.~Wei, and D.~W.~K.~Ng, ``Networked integrated sensing and communications for 6G wireless systems,'' \textit{IEEE Internet Things J.}, vol.~11, no.~17, pp.~29062--29075, Sep.~2024.

\bibitem{ContextSecurity6G} 
A.~Chorti, A.~N.~Barreto, S.~Köpsell, M.~Zoli, M.~Chafii, P.~Sehier, G.~Fettweis, and H.~V.~Poor, ``Context-aware security for 6G wireless: The role of physical layer security,'' \textit{IEEE Commun. Stand. Mag.}, vol.~6, no.~1, pp.~102--108, Mar.~2022.

\end{thebibliography}
\end{document}